\documentclass[structabstract]{aa}  
\usepackage{graphicx}
\usepackage{txfonts}
\def\simlt{\mathrel{\rlap{\lower 3pt\hbox{$\sim$}}\raise 2.0pt\hbox{$<$}}}
\def\simgt{\mathrel{\rlap{\lower 3pt\hbox{$\sim$}} \raise 2.0pt\hbox{$>$}}}
\def\Msun{M_{\odot}}
\def\Zsun{Z_{\odot}}

\begin{document}

\title{The afterglow and host galaxy of GRB\,090205: evidence for a Ly$-\alpha$ emitter at $z=4.65$\thanks{The results
reported in this paper are based on observations carried out at ESO telescopes under programmes Id 082.A-031 and
283.D-5033.}}

\author{P. D'Avanzo\inst{1} \and M. Perri\inst{2} \and D. Fugazza\inst{1} \and 
R. Salvaterra\inst{3} \and G. Chincarini\inst{1,4} \and R. Margutti\inst{1,4} \and X. F. Wu\inst{5,6,7} \and 
C. C. Th\"one\inst{1} \and A. Fern\'andez-Soto\inst{8} \and T. N. Ukwatta\inst{9,10} \and D. N. Burrows\inst{5} \and N. Gehrels\inst{10} \and 
P. Meszaros\inst{5,6,11} \and K. Toma\inst{5,6} \and B. Zhang\inst{12} \and S. Covino\inst{1} \and S. Campana\inst{1} \and V. D'Elia\inst{2,13} \and M. Della Valle\inst{14,15} \and S. Piranomonte\inst{13}}

\offprints{P. D'Avanzo, paolo.davanzo@brera.inaf.it}

\institute{
INAF-Osservatorio Astronomico di Brera, via Bianchi 46, I-23807, Merate, Italy. \and
ASI Science Data Center, via Galileo  Galilei, I-00044, Frascati, Italy. \and
Universit\`a degli Studi dell'Insubria, Dipartimento di Fisica e Matematica, via Valleggio 11, I-22100, Como, Italy. \and
Universit\`a degli Studi di Milano-Bicocca, Dipartimento di Fisica, piazza delle Scienze 3, I-20126, Milano, Italy. \and
Department of Astronomy and Astrophysics, Pensylvania State University, 525 Davey Lab, University Park, PA 16802, USA. \and
Center for Particle Astrophysics, Pensylvania State University. \and
Purple Mountain Observatory, Chinese Academy of Sciences, Nanjing 210008, China. \and
Instituto de F\'isica de Cantabria (CSIC-UC), 39005, Santander, Spain. \and
The George Washington University, Washington, D.C., 20052, USA. \and
NASA/Goddard Space Flight Center, Greenbelt, MD 20771, USA. \and
Department of Physics, Pensylvania State University. \and
Department of Physics and Astronomy, University of Nevada, Las Vegas, NV 89154, USA. \and
INAF-Osservatorio Astronomico di Roma, via di Frascati 33, I-00040, Monteporzio Catone (Roma), Italy. \and
INAF-Osservatorio Astronomico di Capodimonte, salita Moiariello 16, I-80131 Napoli, Italy. \and
International Center for Relativistic Astrophysics, piazza della Repubblica 10, I-65122, Pescara, Italy.
}

    \date{Received; accepted}
 
\abstract
   {}
   {Gamma-ray bursts have been proved to be detectable up to distances much larger than 
   any other astrophysical object, providing the most effective way, complementary to ordinary
   surveys, to study the high redshift universe. To this end, we present here the results of an observational campaign 
   devoted to the study of the high$-z$ GRB\,090205.}
   {We carried out optical/NIR spectroscopy and imaging of GRB\,090205
   with the ESO-VLT starting from hours after the event up to 
   several days later to detect the host galaxy. We compared the results
   obtained from our optical/NIR observations with the available {\it
   Swift\rm} high-energy data of this burst.}
   {Our observational campaign led to the detection of the optical
   afterglow and host galaxy of GRB\,090205 and to the first measure
   of its redshift, $z=4.65$. Similar to other, recent high-$z$ GRBs,
   GRB\,090205 has a short duration in the rest-frame with
   $T_{90,rf}=1.6$ s, which suggests the possibility that it might belong to the short GRBs class. 
   The X-ray afterglow of GRB\,090205 shows a complex and
   interesting behaviour with a possible rebrightening at 500-1000s
   from the trigger time and late flaring activity.
   Photometric observations of the GRB\,090205 host galaxy argue in
   favor of a starburst galaxy with a stellar population younger than
   $\sim 150$ Myr. Moreover, the metallicity of $Z > 0.27 \, Z_{\odot}$ derived
   from the GRB afterglow spectrum is among the highest derived from GRB
   afterglow measurement at high-$z$, suggesting that the burst occured
   in a rather enriched envirorment. Finally, a detailed analysis of the
   afterglow spectrum shows the existence of a line corresponding to 
   Lyman-$\alpha$ emission at the redshift of the 
   burst. GRB\,090205 is thus hosted in a typical Lyman-$\alpha$ emitter (LAE) at
   $z=4.65$. This makes the GRB\,090205 host the farthest GRB host galaxy, spectroscopically confirmed, 
   detected to date.}
  % conclusions heading (optional), leave it empty if necessary {}
   {}


   \maketitle

\section{Introduction}

Gamma-Ray Bursts (GRBs) are powerful flashes of high-energy photons occuring
at an average rate of a few per day throughout the Universe. Thanks to their
optical brightness that typically overshines the luminosity of their host galaxy, they 
are detectable up to extremely high redshift, as clearly 
shown by the recent detection of GRB~090423 at $z\sim 8.2$ (Salvaterra et al. 
2009; Tanvir et al. 2009). This has strengthened the idea that GRBs can be used
as a tool to study the Universe up to (and beyond) the reionization epoch. 
Indeed, GRBs can be used to identify high-$z$ galaxies and study their metal 
and dust content through the identification of metal absorption lines in their
optical afterglow. 

Two classes of GRBs, short and long, have been 
identified on the bases of their observed duration (shorter or longer than $\sim 2$ s)
and spectral hardness (Kouveliotou et al. 1993). In the last years,
observations by the {\it Swift} satellite has questioned this simple 
scheme calling for a classification invoking multiple observational
criteria (see Zhang et al. 2009). 
To this end, prompt emission properties like the isotropic gamma-ray energy release 
($E_{\gamma,iso}$) and the peak energy ($E_{p}$) seem to provide 
a promising tool for GRB's classification, as shown by the $E_{p,i}-E_{\gamma,iso}$ 
correlation (Amati et al. 2008) and its derivations (see, e.g., Lv et al. 2010 and references therein).
While it is widely believed that the majority of long GRBs originate
from the collapse of massive stars, the nature of the progenitors of short ones
is still unclear, though likely linked to the merger of two compact objects.
Long GRBs are typically found to be hosted in low-mass, blue 
galaxies with high specific star formation rates (SSFR), whereas short GRBs 
are generally hosted in more heterogeneous types of galaxies, at least some with lower 
SSFR (see e.g. \cite{Fruchter2006}; \cite{Berger2009}; \cite{Savaglio2009}; \cite{Fong2010}).

In this paper we report the detection of GRB~090205 at $z=4.65$ and 
the study of the properties of its host galaxy, a young starburst. 
The paper is organized as follows. In Section~2, we report the
detection of GRB~090205 by {\it Swift} (Section~2.1) and the discovery and
study of its optical afterglow and of its host
galaxy (Section~2.2). The discussion about the nature of the burst is
given in Section~3.1, whereas the interpretation of its afterglow
is reported in Section~3.2. In Section~3.3, we discuss the nature of the
host galaxy of GRB~090502 and finally, we summarize briefly our main conclusions in
Section~4.

The standard cosmological parameters ($h=0.71$, $\Omega_m=0.27$, 
$\Omega_\Lambda=0.73$) have been assumed and magnitudes are given in the AB
system. All errors are at the $90\%$ confidence level, unless stated
otherwise.

\section{Observations}

\subsection{Swift observation}

GRB~090205 triggered {\it Swift}-BAT (Perri et al. 2009)
on Feb. 2009, 5$^{\rm th}$ at 23:03:14 UT (hereafter, $T_0$). 
The mask-weighted light curve shows a single peak starting at $T_0-5$ s,
peaking at $T_0+3$ s, and returning to background at $T_0+100$ s (Fig.~1).
The duration of the prompt emission is $T_{90}=8.8\pm1.8$ s in the 15-150
keV band.  The time-averaged spectrum from 
 $T_0-2.9$ s to $T_0+6.6$ s in the 15-150 keV band can be fit by a simple
power-law model with photon index $\Gamma=2.15\pm 0.23$. Alternatively,
an equally good fit can be obtained by a cut-off power-law model with 
photon index $a=0.8\pm 1.3$ and observed peak energy $E_p=34 \pm 15$ keV.
A peak energy of $\sim 30$ keV is also found from the relation between
$E_p$ and $\Gamma$ obtained by Sakamoto et al. (2009). The fluence in 
the 15--150 keV band is $F_\gamma=(1.9\pm 0.3)\times 10^{-7}$ erg cm$^{-2}$ 
and the 1-s peak photon flux measured from $T_0+4.09$ s in the 15-150 keV band
is $P=0.5\pm0.1$ ph cm$^{-2}$ s$^{-1}$ (Cummings et al. 2009). 
Following the method described in Ukwatta et al. (2010), we performed the spectral 
lag analysis of the BAT data from $T_0-20$ s to $T_0+20$ s in four energy bands 
(12--25 keV, 25--50 keV, 50--100 keV, 100--350 keV) with a time bin of 1024 ms. 
All lags are consistent with zero, but with relatively 
large uncertainties, given the faintness of the prompt emission. 
{\it Swift}-XRT began to observe the field of GRB~090205 at $\sim 89$ s
after the trigger, identifying a fading uncatalogued
X-ray source located at the UVOT-enhanced position RA (J2000): 14h 43m 38.69s and
Dec (J2000): -27d 51$^\prime$ 09.6$^{\prime\prime}$ with an uncertainty
of 1.8$^{\prime\prime}$ (radius, 90\% confidence, Evans et al. 2009).
{\it Swift}-UVOT began settled observations of the field of GRB~090205 92 s 
after the BAT trigger, but no source was identified at the enhanced Swift XRT 
position.
The burst was declared a ``burst of interest'' by Gehrels \& Perri (2009).

\begin{figure}\label{fig:batlc}
\center{{
\includegraphics[width=\columnwidth]{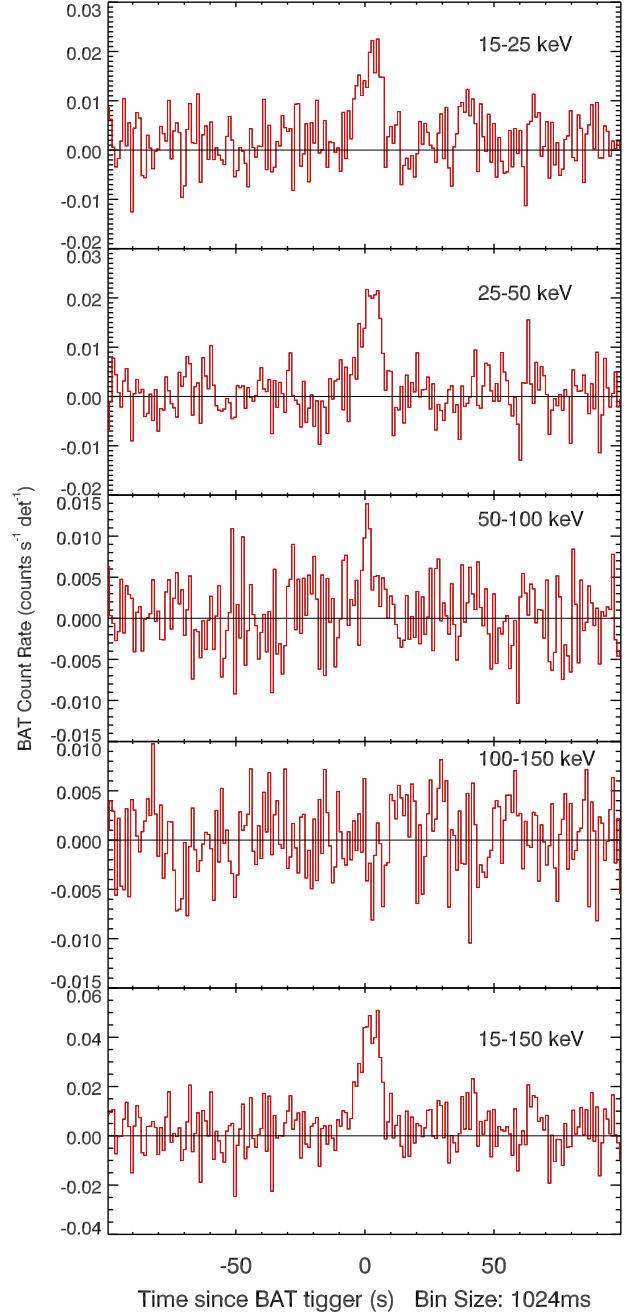}
}}
\caption{Four channels and combined BAT mask-weighted light curve of GRB\,090205. Bin size is 1024 ms.}
\end{figure}

\subsubsection{XRT temporal and spectral analysis}

The XRT data were processed with the XRTDAS software package (v.2.5.0)
developed at the ASI Science Data Center (ASDC) and distributed by HEASARC
within the HEASOFT package (v. 6.7). Event files were calibrated and 
cleaned with standard filtering criteria with the {\it xrtpipeline} task 
using the latest calibration files available in the Swift CALDB.  
The X-ray light curve (Fig.~2) shows a complex behavior. At $\sim T_0+500$s, 
we note a possible re-brightening, while flaring activity is present at 
$\sim T_0+6$ ks and $\sim T_0+20$ ks, respectively.
A fit with a double broken power-law $(F(t)\propto t^{-\alpha})$  gives
indices $\alpha_1=1.36^{+0.37}_{-0.34}$ for $t<T_0+t_{b,1}$, 
$\alpha_2=-0.67^{+1.06}_{-0.66}$ for $T_0+t_{b,1}<t<T_0+t_{b,2}$, and 
$\alpha_3=1.15^{+0.09}_{-0.07}$ for $t>T_0+t_{b,2}$ (excluding the flaring activity), 
where $t_{b,1}=470^{+62}_{-82}$s and $t_{b,2}=1039^{+245}_{-206}$s.
We performed a time resolved spectral analysis of the
X-ray afterglow during the first Swift orbit (spanning from $T_0+100$s to 
$T_0+2$ ks) in three different time 
intervals: (i) $t<470$s (initial decay), (ii) $470$s$<1039$s (rise phase of 
the re-brightening episode), and (iii) $1039$s$<t<2000$s (decay phase of the 
re-brightening episode). For all the three intervals the X-ray spectrum is 
well fitted by an absorbed power-law with photon index $\Gamma_X\sim 2.0$ 
($\Gamma_{X,1} = 1.84 \pm 0.23, \Gamma_{X,2} = 2.00 \pm 0.24, \Gamma_{X,3} = 2.14 \pm 0.20$)
and Galactic $N_H\sim 8\times 10^{20}$ cm$^{-2}$, i.e. no spectral evolution is 
observed during the first orbit, although the data are compatible with a gradual softening of the spectrum. 
No evidence of intrinsic absorption at the redshift of the burst is found. 
The 2(3)-$\sigma$ upper limit is $N_{H,z}<2.3(3.5)\times 10^{22}$ cm$^{-2}$. These 
limits are obtained using the entire XRT dataset (i.e. using $T_0+100$ s -- $T_0+83$ 
ks data).
The time resolved spectral analysis of the X-ray flaring activity has been also 
obtained. The first time interval showing variability (from $T_0+5179$s to $T_0+7696$s) is well fitted by an absorbed power-law 
with photon index $\Gamma_{X} = 2.38 \pm 0.21$ and Galactic $N_H\sim 8\times 10^{20}$ 
cm$^{-2}$, while for the second time interval (from $T_0+16782$s to $T_0+25067$s),
there is some evidence of an harder spectrum with $\Gamma_{X} = 1.56 \pm 0.35$. 
We note, however, that during this flaring activity the photon index values are
consistent (at the 90\% CL) with the values measured at earlier times.

\begin{figure}\label{fig:xlc}
\center{{
\includegraphics[height=8cm,angle=-90]{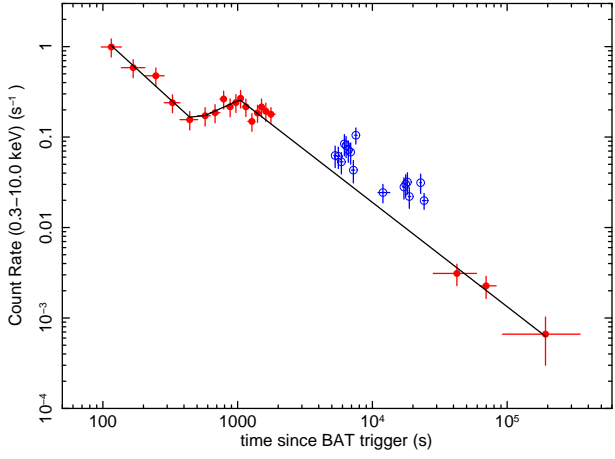}
}}
\caption{X-ray light curve in count rate. The solid line shows the best power-law fit obtained excluding the flaring activity (marked by open circles) present at $\sim T_0+6$ ks and $\sim T_0+20$ ks, respectively. The conversion factor is 1 cts s$^{-1}$ = $3.6 \times 10^{-11}$ erg cm$^{-2}$ s$^{-1}$.  Errors are at 68\% c.l.}
\end{figure}

\subsection{Optical/NIR observations}

A complete log of all our optical/NIR ground based observations is reported in
Tab.~\ref{tab:log}.

\begin{table*}
\caption{\label{t7}VLT observation log for GRB\,090205.}
\centering
\begin{tabular}{ccccccc} \hline
Mean time          &  Exposure time             & Time since GRB & Seeing    &  Instrument  &  Magnitude         & Filter \\
(UT)               &  (s)                       & (days)         & (\arcsec) &              &                    &	  \\ \hline
2009 Feb 06.25042  &  $2  \times 180$           &  00.28984      & 0.7       & VLT/FORS1    & $22.26 \pm 0.04$	 & $R$    \\
2009 Feb 06.25751  &  $2  \times 180$           &  00.29693      & 0.7       & VLT/FORS1    & $20.80 \pm 0.02$	 & $I$    \\
2009 Feb 07.30501  &  $4  \times 180$           &  01.34443      & 1.0       & VLT/FORS1    & $24.51 \pm 0.18$	 & $R$    \\
2009 Feb 26.29729  &  $7  \times 180$           &  20.33671      & 0.9       & VLT/FORS1    & $26.40 \pm 0.26$	 & $R$    \\
2009 Mar 11.36257  &  $40 \times 120$           &  33.40199      & 0.7       & VLT/FORS1    & $-$	         & $I$    \\
2009 Mar 24.22352  &  $20 \times 180$           &  46.26294      & 0.7       & VLT/FORS1    & $25.22 \pm 0.13$	 & $I$    \\
2009 Mar 28.42134  &  $2 \times 30 \times 18$   &  50.46076      & 0.8       & VLT/ISAAC    & $> 23.3$	         & $J$    \\
2009 Aug 10.02094  &  $4 \times 15 \times 64$   & 185.06036      & 0.9       & VLT/HAWKI    & $> 24.4$	         & $J$    \\
2009 Aug 10.52803  &  $4 \times 15 \times 65$   & 185.56745      & 0.8       & VLT/HAWKI    & $> 23.9$	         & $Ks$   \\ 
2009 Aug 11.02510  &  $4 \times 15 \times 65$   & 186.06452      & 0.6       & VLT/HAWKI    & $> 24.2$	         & $H$    \\ \hline
2009 Feb 06.36422  &  $1 \times 1200$           &  00.40364      & 0.7       & VLT/FORS1    & $-$	         & $300V+GG375$ \\ \hline 
\end{tabular}
\tablefoot{
Magnitudes are in the AB system and are not corrected for
Galactic absorption. Errors and upper limits are given at $1\sigma$ and $3\sigma$ confidence level respectively.
}
\label{tab:log}
\end{table*}

\subsubsection{Optical/NIR imaging}

\begin{figure}\label{fig:lco}
\center{{
\includegraphics[height=8cm,angle=-90]{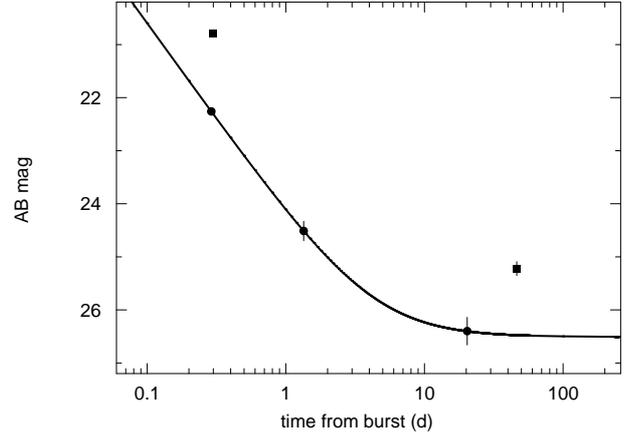}
}}
\caption{Optical $R-$ ({\it dots}) and $I-$band ({\it squares}) afterglow light curve. The solid line shows the best power-law fit. Magnitude are in AB system and are not corrected for Galactic extinction. Errors are at 68\% c.l.}
\end{figure}

We observed the field of GRB~090205 with the 
ESO-VLT in imaging mode starting about 7.1 hours after the burst. 
Observations were carried out in $R$ and $I-$band with the FORS1 camera.
Within the enhanced X-ray position, we identified a source at the following 
coordinates: RA(J2000)=14h 43m 38.70s and Dec(J2000)=-27d 51$^\prime$ 10.0$^{\prime\prime}$ with an
uncertainty of 0.3$^{\prime\prime}$ (D'Avanzo et al. 2009a). 
The source is detected in both bands with $R_{AB}=22.26\pm0.04$
and $I_{AB}=20.80\pm 0.02$, the red color suggesting a high redshift 
object. All values given here are not corrected for Galactic extinction of 
$E(B-V)=0.117$ (Schlegel et al. 1998).  This result has been confirmed by almost contemporary GROND 
observations (Kruehler \& Greiner 2009): the object is very well detected
in the $i^\prime$-band, marginally in the $r^\prime$-band and not in the 
$g^\prime$-band. Interpreting the non detection in the $g^\prime$-band
and the large $r^\prime-i^\prime$ color as due to Lyman-$\alpha$ absorption in the GRB host, 
a photometric redshift of $4.7\pm0.3$ has been derived.

We continued to follow the fading of the GRB afterglow with the FORS1
camera on VLT. A second epoch image was obtained $\sim 1.34$ days after the trigger
in the $R-$band (D'Avanzo et al. 2009b). The source is detected with $R_{AB}=24.51\pm0.18$. The
fading behaviour confirms it as the optical afterglow of GRB~090205.
Late time, multiband observations were also carried out to search for the 
GRB host galaxy, showing a flattening in the $R$ and, possibly, in the
$I-$band light curves (see Sect~\ref{sec:host}). 
Assuming a power-law decay $F(t) = F_0 + kt^{-\alpha}$, the decay index is
$\alpha_{opt} = 1.35^{+0.26}_{-0.19}$, steeper (but consistent within the errors) than
the decay index derived from the X-ray light curve at the same epochs.
A plot of our $R-$ and $I-$band observations is shown in Fig.~3.

\subsubsection{Spectroscopic Observations}

We observed the source with the ESO VLT about 9.0 hours after the
burst with the FORS1 camera in spectroscopic mode. We took a 20 min
spectrum with the 300V grism (11~\AA{} FWHM) using a slit with $1''$ width. 
We covered wavelength range 4000-10000 \AA$\;$with a resolution of
R=440 (Th\"one et al. 2009; Fugazza et al. 2009).  The spectrum was reduced 
with standard tasks in IRAF, combined and flux calibrated using observations 
of the standard star EG274 taken on Feb. 7, 2009.

The most prominent feature in the
spectrum is the Damped Lyman Alpha system (DLA) at 6873 \AA{} in the observer
frame. Furthermore, we detect Ly$-\beta$ and Ly$-\gamma$ in absorption as well as
the Lyman break. Redwards of the DLA, we detect a range of absorption
lines from the host galaxy, SII, SiII, CII, SiIV and CIV (see
Table \ref{tab:lines}). We also detect the fine structure transition of
SiII* $\lambda$\,1264\,\AA{}, but no other fine structure lines could
be identified.  From
those lines, we determine the redshift of the GRB to z$=4.6503\pm0.0025$. 
A plot of the combined spectrum with the lines identified is shown in Fig.~4.
A detailed analysis of the spectrum shows the presence of
an emission line at $\sim 6873$\,\AA{} (Fig.~5). 
At the redshift of the GRB, it corresponds to Ly$-\alpha$ emission at 
1215\,\AA. We will discuss it more in detail in the next section.

We fitted the red wing of the DLA using the MIDAS fitlyman
package. The resulting column density was determined to be log
N$_H$/cm$^{-2}=$20.73$\pm$0.05. The HI column density lies below 
the average neutral hydrogen column density for GRB-DLAs of log
N$_H$/cm$^{-2}=$21.6 (Jakobsson et al. 2006; Fynbo et al. 2009). 
This is in agreement with the simulations of Nagamine et al. (2008)  
which show that the mean DLA column density decreases with increasing redshift. 
On the other hand, the relatively low number of GRBs at redshift $\geq 3-4$ 
with measured HI column density and the probable observational bias against the most dusty environments (Jakobsson et al. 2006; Fynbo et al. 2009) 
do not enable yet to firmly check the existence of this anti-correlation.

The majority of the absorption lines detected are saturated and
therefore do not allow a reliable determination of the column
density. Mildly saturated lines, which we define here as lines
with an EW of $<$ 0.5\,\AA{}, do not lie on the linear part of the 
curve of growth and the derived column densities can only be considered 
as lower limits. For SII, SiII* and OI (EWs from 0.30 to 0.49, Tab.~2) we 
obtain, assuming a linear relation between the EW and the column density, 
log N/cm$^{-2}=$15.3, log N/cm$^{-2}=$13.4 and log N/cm$^{-2}=$14.8, respectively.
We take the column density limit derived from SII to determine the metallicity of the
host as SII is not affected by depletion onto dust. Using the solar abundances 
reported in Asplund et al. (2009), we find the metallicity in the host along the line of 
sight to be $[M/H] > -0.57$ or $Z>0.27\;\Zsun$.

\begin{table}[h]            
\caption{\label{t7}Measured wavelengths, derived redshift and equivalent widths (EWs) of detected absorption lines}
\centering                        
\begin{tabular}{l l l l l l}       
\hline\hline
$\lambda_\mathrm{obs}$ & $\lambda_{rest}$&id& z & EW$_\mathrm{rest}$& log N\\
 $[\AA{}]$ & $[\AA{}] $& &  &[\AA{}] & [cm$^{-2}$]\\
\hline
5495	&972.54	 & Ly$\gamma$ & ---	& ---	& ---\\
5791	&1025.72 & Ly$\beta$ 	& ---	& ---	& ---\\
6870  	&1215.67 & Ly$-\alpha$	& ---	& 98.5	& 20.73$\pm$0.05\\
7091.20	& 1253.81 & {S}{II} & 4.6557 & 0.30 &$>$15.3\\
7120.35 & 1259.52 & {S}{II} & --- & 1.38$\pm$0.04 & ---\\
--- & 1260.42 & {Si}{II} & --- & {\it blended} & --- \\
--- & 1260.53 & {Fe}{II}& --- & {\it blended} & ---\\
7149.40	&1264.74 & {Si}{II*} & 4.6529 & 0.42$\pm$0.03 & $>$13.4\\	
7357.65  &1302.17 & {O}{I} & 4.6503 &  0.49$\pm$0.02 & $>$14.8\\
7372.76	&1304.37 & {Si}{II} & 4.6523 & 0.95$\pm$0.04 & ---\\
7543.54 & 1334.53 & {C}{II} & --- & 1.96$\pm$0.50 &  ---\\
--- & 1335.71 & {C}{II}	 & --- & {\it blended}	& ---\\
7873.45 &1393.76 & {Si}{IV} & 4.6490 & 2.07$\pm$0.16 & ---\\
7919.98  &1402.77 & {Si}{IV} & 4.6460 & 3.91$\pm$0.51 & ---\\
8627.73  &1526.71 & {Si}{II} & 4.6512 & 1.36$\pm$0.04 & ---\\
8754.28  & 1550.78 &{C}{IV}& 4.6450 &2.89$\pm$0.04& ---\\
\hline \hline
\end{tabular}
\tablefoot{The EW for the blended systems
include the contributions from all transitions in the blended
line. For blended systems, the redshift is not mentioned due to the
large uncertainty. We only give lower limits on the column density for
mildly saturated lines which we define here as EW $<$ 0.5 \AA{}
in the restframe.
}
\label{tab:lines}
\end{table}

\begin{figure*}\label{fig:spectrum}
\center{{
\includegraphics[height=9cm]{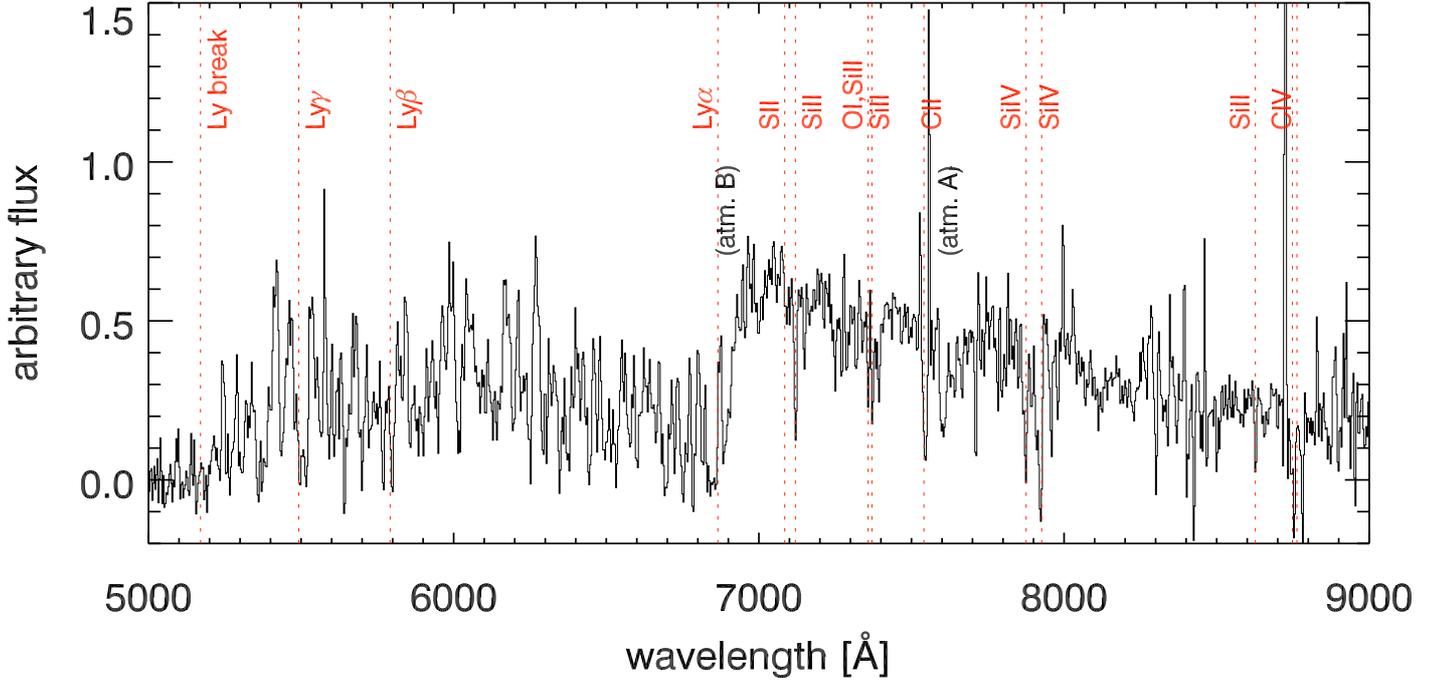}
}}
\caption{VLT/FORS1 spectrum of the GRB~090205 afterglow at $z~=4.6503~\pm~0.0025$.}
\end{figure*}

\subsubsection{Host galaxy observations}\label{sec:host}

We continued to monitor the field of GRB~090205 at late times to further study
the GRB host. We obtained an image in the $R-$band with the FORS1 camera $\sim 20.3$ days 
after the trigger. A faint object at a magnitude of $R_{AB} \sim 26.4 \pm 0.3$  
has been identified very close to the position of the GRB afterglow. This detection represents a 
flattening in the $R-$band light curve, that we interpret as due to the host galaxy (Fig.~3).
Further $I-$band monitoring, carried out $\sim 46.3$ days after burst\footnote{As reported in
Tab.~\ref{tab:log}, another epoch of $I-$band imaging was taken at $t-T_0 \sim 33$ d but, 
erroneously, no dithering was performed among different frames. The resulting image was thus 
highly affected by fringing (see http://www.eso.org/sci/facilities/paranal/instruments/fors/) 
and it was not possible to obtain reliable measures of photometry.} reveals an object with 
$I_{AB} = 25.2 \pm 0.1$. Comparing this detection with the previous one obtained in the $R-$band,
the resulting unabsorbed $R-I$ color is consistent with that of the 
afterglow. This suggests a flattening of the light curve in the $I-$band too, in agreement with
the hypothesis that we detected the host galaxy of GRB\,090205.
We also carried out deep, late-time ($t-T_0 \sim 180$ d) NIR observations of the field of 
GRB~090205 with VLT/HAWK-I in $JHK-$bands. The host is not detected in any
of the observed bands up to a limiting AB magnitude of $J>24.4$, $H>24.2$ and 
$Ks>23.9$ ($3\sigma$ c.l.).
The results are reported in Tab.~\ref{tab:log} and in Figs.~6,7.

As already mentioned, the afterglow spectrum shows  an emission line at 
$\sim 6873$\,\AA{ } superposed on the Ly-$\alpha$ absorption, 
corresponding to Ly-$\alpha$ emission at the same redshift of the GRB. In
order to check the reliability of the line detection, and to exclude the possibility that it 
is due to some atmospheric emission or absorption contaminating feature, we performed a
detailed analysis of the 2-D spectrum (see Fig.~5). At the
wavelength corresponding to the Ly$-\alpha$ line emission we measure $2101\pm 51$
counts (sky+object). The counts corresponding only to sky are $1836\pm
21$, so that the object counts are $265\pm55$ ($68\%$ c.l.). The corresponding
signal-to-noise ratio is 5.2. 
Another striking evidence we obtain from the 2-D spectrum is the measure 
of a spatial displacement of $1.3 \pm 0.9$ pixels (equivalent to $0.3'' \pm 0.2''$) from the centroid of the 
afterglow continuum trace and the ``spot'' corresponding to the Ly$-\alpha$ 
emission (see Fig.~5). 
Doing precise astrometry on our afterglow and host galaxy images obtained with FORS1, 
we measure the same offset between the afterglow and the host galaxy positions 
($0.4'' \pm 0.3''$, corresponding to a physical offset of about 3 kpc), thus 
making stronger the hypothesis that this emission line is 
really due to Ly$-\alpha$ from the host galaxy. 
Using the flux-calibrated afterglow spectrum we derive a flux of
$1.82\times 10^{-17}$ erg s$^{-1}$ cm$^{-2}$. This
flux transforms into a Ly$-\alpha$ luminosity of $4.27\times
10^{42}$ erg s$^{-1}$. We note that this value is in the range of 
luminosities observed for the other GRB-LAE hosts\footnote{GRB~971214 ($z=3.42$), GRB~000926 ($z=2.04$), GRB~011211 ($z=2.14$), GRB~021004 ($z=2.33$), GRB~030323 ($z=2.66$), and GRB~030429 ($z=2.66$).}, i.e. $1-5\times 10^{42}$ erg s$^{-1}$ (Jakobsson et al. 2005).

\begin{figure}\label{fig:Lya}
\center{{
\includegraphics[width=8cm]{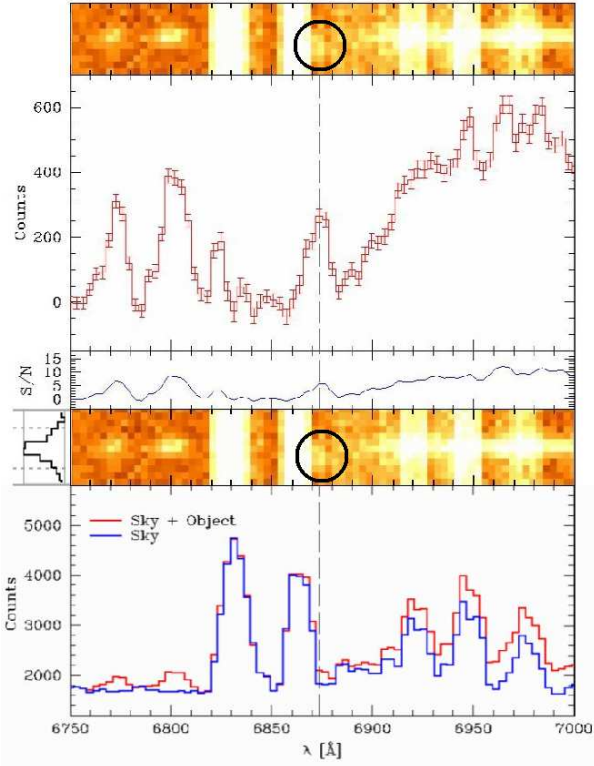}
}}
\caption{Detailed analysis of the Ly$-\alpha$ emission line detection. The
top panel shows the blow up of the region centered on the Ly$-\alpha$ emission
at 6873.45\AA\, (dashed line), corresponding to 1215.67\AA\, in the emitter rest
frame. The central panel shows the corresponding signal-to-noise ratio.
The bottom panel reports the counts from the sky (blue line) and for
the sky+object (red line). The position of the Ly$-\alpha$ emission in the 
2-D spectrum (shown in both panels), is marked by a circle and 
has an offset with respect to the afterglow continuum 
(corresponding to $0.3''$; see Sec. 2.2.3 for details).}
\end{figure}

\section{Discussion}

\subsection{Burst classification}

One interesting aspect of this burst is that, similarly to other, high-$z$
GRBs (e.g. GRB~080913 at $z=6.7$, Greiner et al. 2009; GRB~090423 at $z=8.2$,
Salvaterra et al. 2009, Tanvir et al. 2009), it shows a short duration in 
the emitter rest frame, $T_{90,rf}\sim 1.6$ s. 
A short rest frame duration was recently proposed as a possible indicator (among others) of GRBs originated from a compact-star-merger progenitor (or Type I GRBs; Zhang et al. 2009). While the spectral lag analysis is
inconclusive regarding the nature of this burst, owing to the faintness of 
the prompt emission (Sec.~2.1), the BAT spectrum appears to be softer with respect to typical short GRBs. 
At $z=4.65$, the isotropic gamma-ray energy release in the redshifted 15-150 keV band is 
$E_{\gamma,iso}=7.86\pm 1.21\times 10^{51}$ erg and the intrinsic peak
energy is $E_{p,i}=192\pm 85$ keV. These values make GRB~090205 consistent 
with the observed  $E_{p,i}-E_{\gamma,iso}$ correlation (Amati et al. 2008), that is
known to be followed only by long GRBs (see also Piranomonte et al. 2008) 
and proposed as an indicator of GRBs with a massive stellar collapse origin (Type II GRBs; Zhang et al. 2009). Indeed, the $E_{p,i}-E_{\gamma,iso}$ 
correlation has been used recently to support the long classification of 
a few rest-frame short duration bursts such as GRB~090423 
(Salvaterra et al. 2009) and GRB~090426 (Antonelli
et al. 2009). A Type II classification for GRB~090205 is also 
supported by applying the classification method reported in Lv et al. (2010), 
being $\epsilon = E_{\gamma,iso,52}/{E_{p,z,2}}^{5/3}\sim 0.26$, where 
$E_{\gamma,iso,52} = E_{\gamma,iso}/10^{52}$ erg and $E_{p,z,2}=E_p(1+z)/10^2$ keV. 
This value puts GRB~090205 in the high-$\epsilon$ 
regime, which is related to long (Type II) GRBs.

In conclusion, even if a massive stellar collapse origin for GRB~090205 may appear 
puzzling (although not unheard of)\footnote{Despite their short rest-frame duration, the high$-z$ GRB\,080913 and GRB\,090423 were classified as Type II bursts (Zhang et al. 2009).} in light of its rest frame short duration, the 
prompt emission properties of this GRB favors for a Type II 
classification. Furthermore, we note that, while the existence at high-$z$ of a population 
of bursts originating from the merging of double compact objects is expected on 
theoretical ground (Belczynski et al. 2010), their detection would imply a 
very flat luminosity function for the short burst population, in contrast 
with a recent analysis of  BATSE and {\it Swift} 
data (Salvaterra et al. 2008).

\subsection{Afterglow theoretical interpretation}

As shown in Sect.~2.1, the X-ray afterglow evolution can be divided into
three stages. The closure relation for the first stage is
$\alpha_1 - 1.5 \beta_1 = 0.10^{+0.51}_{-0.48}$, which can not
constrain any model due to the large scatter of the error bars.
The closure relation for the third stage is
$\alpha_3 - 1.5 \beta_3 = -0.56 \pm 0.31$, quite consistent with the
theoretical expectation of $\alpha-1.5\beta=-0.5$, where
$\alpha=(3p-2)/4$ and $\beta=p/2$. Therefore, for $t>10^3$ s, the
inferred power law index of electron energy distribution shaped by
shock acceleration is $p = 2.20^{+0.12}_{-0.09}$.
Therefore, the simplest forward shock model for the first and
third stages corresponds to the X-ray band being above the cooling
and typical frequencies of synchrotron radiation, i.e., $\nu_x > max (\nu_m,\nu_c)$.

The second stage shows a rise of the X-ray flux with time. Interpreting it 
as due to the emergence of the Synchrotron-Self-Compton component in 
the X-ray band, there should be significant spectral hardening around the
transition time $t\sim 500$ s, which is contrary to what we
observed. Alternatively, the rise may be due to the
continuous energy injection ($L_{\rm {inj}}\propto t^{-q}$) from
late time central engine activities (e.g., Dai \& Lu 1998) 
or refreshed shock (Rees \& Meszaros 1992) by a late
time ejecta with varying Lorentz factors within the ejecta
($M(>\Gamma)\propto\Gamma^{-s}$). Since the X-ray spectral index
in the rising phase is steep, the characteristic frequencies
$\nu_c$ and $\nu_m$ should be below the X-ray band. 
According to Table 2 of Zhang et al. (2006), we use the relation of 
$\alpha_2=(q-2)/2+(q+2)\beta_2/2$ and obtain 
$q = 2(1+\alpha_2-\beta_2)/(1+\beta_2) = -0.67^{+1.07}_{-0.68}$, where 
$\alpha_2 = -0.67^{+1.06}_{-0.66}$ and $\beta_2 = 1.00 \pm 0.24$. 
For the matter-dominated injection model, $s = (10-7q)/(2+q) =
11.05^{+9.23}_{-14.52}$ for the ISM case and $s = 4/q - 3 = -8.97^{+6.06}_{-9.53}$ for
the wind case. Since $s<0$ is unphysical, the wind model is
therefore not favored if the rebrightening is due to a
matter-dominated refreshed shock. The afterglow kinetic energy
after the rising phase is increased by a factor of
$\left(\frac{1000\;\rm{s}}{500 \;\rm{s}}\right)^{1-q}\sim
3.2^{+3.5}_{-1.2}$.

The X-ray afterglow clearly shows the presence of late-time temporal
variability ($4\,\rm{ks}<t<20\,\rm{ks}$).  The variable
afterglow is characterised  by a flux contrast $\Delta F/F\sim 3$,  where
$\Delta F$ is the flux enhancement  due to the possible flares and $F$
is the flux level of the underlying continuum.
This together with the upper limit on the variability ratio $\Delta 
t/t<0.3$, places the GRB\,090205 possible X-ray flares at the boundary between
density fluctuations produced by many regions viewed off-axis and
refreshed shocks (see \cite{Ioka2005}; \cite{Chincarini2007}, their
Fig. 15).  However, the low statistics prevents us from drawing quantitative
conclusions on both the temporal (see Chincarini et al., 2010 
for an updated analysis on 113 GRB X-ray flares) and spectral behaviour of 
this possible flaring activity (\cite{Falcone2007}).

The Galactic extinction corrected {\it I}-band flux density at $t \sim
25$ ks is $\sim 21.4$ $\mu$Jy. At this time, the 0.3 - 10 keV count rate is
$\sim 6.3\times 10^{-3}$ counts s$^{-1}$, corresponding to a flux of $2.3 \times
10^{-13}$ erg cm$^{-2}$ s$^{-1}$. Assuming the late-time X-ray spectral
index $\beta_3=1.07$, the X-ray flux density at $\nu_X = 10^{18}$ Hz is
$\sim 6.2 \times 10^{-3}$ $\mu$Jy. So the near infrared to X-ray overall
spectral index at $t \sim 25$ ks is $\beta_{NIR-X} \sim 1.0$, suggesting
that the optical/NIR and the X-ray emission are from the same origin.

In conclusion, the X-ray and optical afterglow can be explained within the
standard forward shock model with $\nu_c, \nu_m <$ $\nu_{\rm opt}$
and $\nu_X$. The early rebrightening in the X-ray afterglow can be
interpreted as due to the energy injection into the forward shock by
the central engine. We note that more complex modelling of the rebrightening
phase (i.e. two-component models) are not strictly required by the data.

\subsection{GRB host}

Our photometric campaign carried out with VLT/FORS, ISAAC and HAWK-I (see Sect.~2.2.3)
allows us to detect the GRB host galaxy in the R and I band and to put strong
upper limits on the continuum in the NIR bands (J, H, and K). The observed 
magnitude and limits are reported in Table~\ref{tab:log} and shown in Figs.~6,7. 
The blue color, $(I-K)_{AB}<1.1$, argues for a starburst galaxy, whereas
ellipticals, Sab, Scd type of galaxy are discarded (see Fig.~6). We 
therefore model the photometric data with a family of synthetic starburst 
SEDs computed from the outputs of the Starburst99 code (Leitherer et al.
1999; Vazquez \& Leitherer 2005). We adopt a Salpeter initial mass function
in the mass range 0.1-100 $\Msun$ and a metallicity of $Z=0.4\;\Zsun$ consistently with the metallicity obtained from the GRB 
afterglow spectrum. Different ages of the stellar population are considered
and the synthetic SEDs are normalized  to reproduce 
the observed magnitude in the I-band. The absorption due to the intergalactic
medium shortwards the Ly$-\alpha$ has
been modelled as in Salvaterra \& Ferrara (2003, see Section~2.2). The theoretical
SEDs are shown in Fig.~7 from top to bottom with stellar ages of 500, 100, 50, 
10 Myr, respectively. We find that the upper limits in the NIR bands provide
a strong limit to the age of the stellar population. In order not to exceed
the J and K band upper limits, the stellar population should be younger than
$\tau<150$ Myr. In this case, the
corresponding stellar mass is $M_\star<5\times 10^{10}\;\Msun$, in agreement with 
average mass of long GRB host galaxies ($10^{8.5}-10^{10.3}\;\Msun$ Savaglio et al. 2009). We neglect
here the possible presence of dust inside the host galaxy. However, we
note that dust extinction would result in a reddening of the host SED, 
strengthening our limits on the stellar age and mass.

As described in Sect.~2.2.3, we found evidence that the host galaxy of GRB\,090205 is a 
Ly$-\alpha$ emitter. 
The Ly$-\alpha$ emission line lies at $z = 4.6537 \pm 0.0014$ that, compared to the redshift measured from the absorption lines (Sec.~2.2.2), gives ${\Delta}z = 0.0034 \pm 0.0029 \, ({\Delta}v = 180 \pm 153$ km s$^{-1})$. This is in line with the results obtained through spectroscopic studies performed on large samples of Lyman break galaxies (LBG) at $z \sim 3$ that show velocity offsets between the Ly$-\alpha$ emission and interstellar absorption line redshifts of the order of $\sim 600$ km s$^{-1}$, with large dispersion ($\sim 300-500$ km s$^{-1}$; see e.g. Adelberger et al. 2003; Shapley et al. 2003; Bielby et al. 2010). Such kinematics is usually interpreted as due to large-scale outflows caused by supernova--driven wind, resulting from intense star formation, that blueshift absorption lines from the interstellar gas. 
At the redshift of the burst, the Ly$-\alpha$ luminosity is
$4.3\times 10^{42}$ erg s$^{-1}$. This value lies in the range of luminosities of other LAEs 
identified by dedicated surveys at $z\sim 4.5$ (Finkelstein et al. 2007; 
Shioya et al. 2009; Wang et al. 2009). 
In particular, Shioya et al. (2009) in a recent survey of $z\sim 4.8$ LAE in 
the COSMOS 2 square  degree field compute the Ly$-\alpha$ luminosity
function of these objects, measuring 
$L_\star=8^{+17}_{-4}\times 10^{42}$ erg s$^{-1}$. Similarly, Wang et al. 
(2009) find $L_\star=6.3\pm 1.5\times 10^{42}$ erg s$^{-1}$ for a 
sample of 110 LAEs detected in the Large Area Lyman Alpha (LALA) survey.
Our findings suggest thus that   
this burst exploded into a 0.6-0.7 $L_\star$ LAE at that redshift. 
Transforming the luminosity in a star-formation rate using the formula from
\cite{Kennicutt98} for H$\alpha$ and assuming a factor of 8 between
H$\alpha$ and Ly$-\alpha$, we derive a SFR of 4.2 M$_\odot$\,yr$^{-1}$
which is among the typical values found for other Ly-$\alpha$ emitters hosting
GRBs (see, e.g. Jakobsson et al. 2005) and typical galaxies hosting GRBs (Savaglio et al. 2009).
However the above values should be interpreted as lower limits. 
During the acquisition of the spectrum, the slit was centered on the afterglow position, 
so that we lost part (about 50\%) of the Ly$-\alpha$ flux coming from the host galaxy, 
due to the $0.4''$ offset we discussed in Sec. 2.2.3. 

\begin{figure}\label{fig:host}
\center{{
\includegraphics[width=8cm]{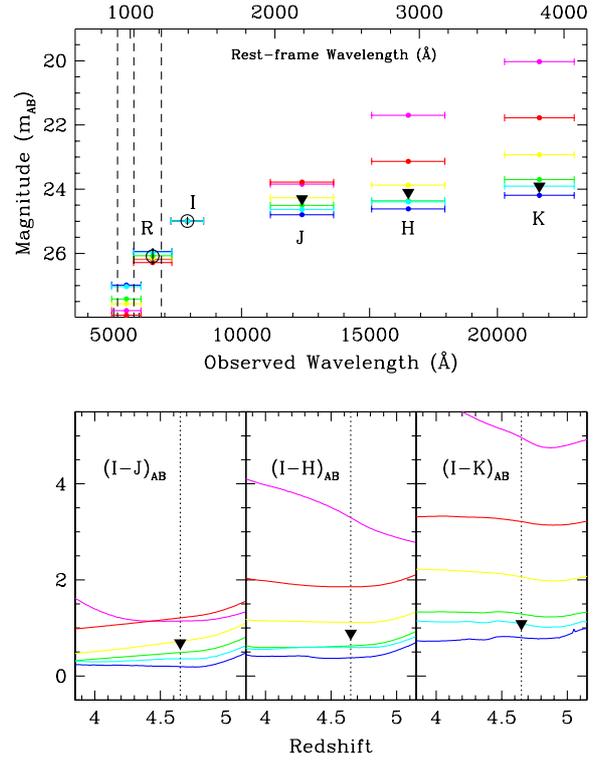}
}}
\caption{Top panel: available data and expected magnitude of
the GRB~090205 host for different galaxy types. The SEDs are normalized to 
reproduce the
$I$ band measure. Open circles (filled triangles) represent the data (upper limits), 
whereas the horizontal lines
show the expected AB apparent magnitude of the object in the $VRIJHK$ bands
for different host galaxy types: magenta is for Elliptical, red
for Sab, yellow for Scd, green for Irregular, cyan and blue for starburst 
galaxies. Vertical lines mark the position of Lyman limit, Ly$\beta$ and Ly$-\alpha$ (from left to right), respectively.  Bottom panels: expected 
$(I-J)$, $(I-H)$ and $(I-K)$ colors for the different galaxy types. Observational limit on the colors are plotted with triangles.
}
\end{figure}

\begin{figure}\label{fig:host}
\center{{
\includegraphics[width=8cm]{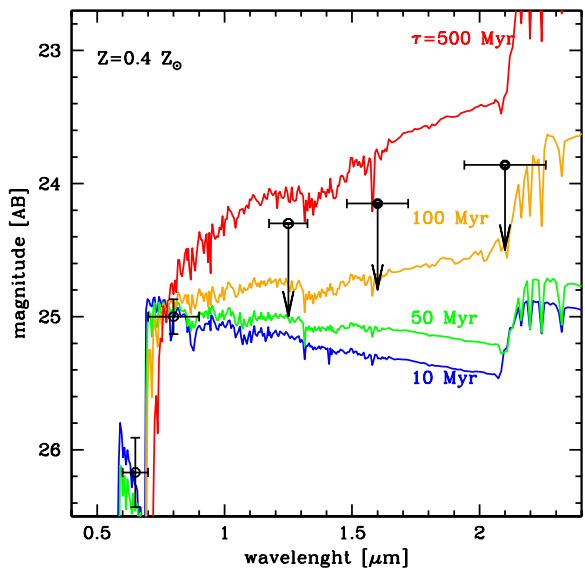}
}}
\caption{Observations of the host galaxy of GRB~090205 shown with data 
points and upper limits are compared with synthetic SEDs for a starburst
galaxy with different stellar ages $\tau$. From top to bottom, lines 
corresponds to 
$\tau=500$, 100, 50, 10 Myr, respectively. We assume a Salpeter IMF between
0.1-100 $\Msun$,  $Z=0.4\;\Zsun$ 
and no dust extinction in the host galaxy. The SED are normalized to reproduce
the observation in the $I$ band. 
}
\end{figure}

Our analysis of the GRB afterglow spectrum provides also a lower limit on the galaxy metallicity,
$Z\ge 0.27\;\Zsun$. Given the limit on the stellar mass obtained above, 
this metallicity is consistent with the mass-metallicity measured
for Lyman Break Galaxies at $z=3-4$ (Mannucci et al. 2009). We caution however 
that the metallicity probed by absorption lines does not necessarily probe the 
metallicity of the entire GRB host, but more likely of the line of sight 
towards the inner, star-forming region of the GRB host. 
The metallicity of GRB~090205 is among the highest
determined for high-$z$ GRBs.  Comparing the metallicity of GRB~090205 to 
those determined for other GRBs at various redshifts,
we find little or no evolution with redshift, in contrast with what 
found for the QSO selected DLA population (Fynbo et al. 2006; Savaglio et al. 2009).

\section{Conclusions}

We report the detection and study of GRB~090205 at $z=4.65$. 
Similar to other, recent high-$z$ GRBs, GRB~090205 has a short duration
in the rest-frame with $T_{90,rf}=1.6$ s. However, the analysis of its prompt emission 
properties favor a massive stellar collapse origin. 
The X-ray afterglow of GRB~090205 shows a complex behaviour with a possible rebrightening at 
500-1000s from the trigger and flaring activity at later times. The X-ray and optical 
afterglow can be explained within the
standard forward shock model with $\nu_c, \nu_m <$ $\nu_{\rm opt}$
and $\nu_X$, where the early rebrightening in the X-ray afterglow can be
interpreted as due to the energy injection into the forward shock by
the central engine.

Finally, we report the detection of the host galaxy of GRB~090205, 
which is found to be a typical LAE at $z = 4.65$, making it the farthest 
GRB host galaxy spectroscopically confirmed. 
The blue color indicates a starburst galaxy with a young ($\tau<150$
Myr) stellar population, further supporting the long classification for this GRB. 
The obtained mass and SFR are in line with typical values of GRB host galaxies, while 
the metallicity derived
from the GRB afterglow spectrum is among the highest derived from GRB afterglow
measurement at high-$z$, suggesting that the burst occured
in a rather enriched envirorment. 

In conclusion, GRB~090205 clearly shows  that GRBs can be used as signpost
of young, starburst galaxies at high-$z$ that are thought
to be the dominant galaxy population at those epochs. 
Thanks to the brightness of their afterglow, metal lines can be
easily identified providing, together with follow-up photometric observation
of their host galaxies, a new way to measure the mass-metallicity relation
and its evolution through cosmic times.

\begin{acknowledgements}
We thank the referee for his/her useful comments and suggestions. 
We acknowledge support by ASI grant SWIFT I/011/07/0. This research has made use of the XRT Data Analysis Software (XRTDAS) 
developed under the responsibility of the ASI Science Data Center (ASDC), Italy. We acknowledge the invaluable help from the ESO staff at Paranal in carrying out our target-of-opportunity observations. 
\end{acknowledgements}

\end{document}